# Yield Enhancement of Digital Microfluidics-Based Biochips Using Space Redundancy and Local Reconfiguration*

Fei Su, Krishnendu Chakrabarty and Vamsee K. Pamula
Department of Electrical & Computer Engineering
Duke University, Durham, NC 27708
E-mail: {*fs, krish, vkp*}@ee.duke.edu

**Abstract**

*As microfluidics-based biochips become more complex, manufacturing yield will have significant influence on production volume and product cost. We propose an interstitial redundancy approach to enhance the yield of biochips that are based on droplet-based microfluidics. In this design method, spare cells are placed in the interstitial sites within the microfluidic array, and they replace neighboring faulty cells via local reconfiguration. The proposed design method is evaluated using a set of concurrent real-life bioassays.*

## 1. Introduction

Microfluidics-based biochips promise to revolutionize clinical diagnostics, DNA sequencing, and other procedures involving molecular biology [1]. In contrast to continuous-flow microfluidic biochips consisting of permanently etched micropumps, microvalves, and microchannels [1], droplet-based microfluidic biochips relying on the manipulation of individual droplets through electrowetting and referred to as "digital microfluidics" have been demonstrated [2]. Digital microfluidics offer dynamic reconfigurability, whereby groups of cells in a microfluidic array can be reconfigured to change their functionality during the concurrent execution of a set of bioassays. These features make digital microfluidics a promising platform for massively-parallel DNA analysis, automated drug discovery, and real-time biomolecular recognition.

Future advances in fabrication technology will allow increased integration of microfluidic components in composite microsystems. The 2003 International Technology Roadmap for Semiconductors (ITRS) anticipates that microfluidic biochips will soon be integrated with electronic components in system-on-chip (SOC) design [3]. It is expected that several bioassays will then be concurrently executed in a single microfluidic array [4]. However, as in the case of integrated circuits, increase in density and area of microfluidics-based biochips will reduce yield, especially for new technology nodes. Low yield is a deterrent to large scale and high-volume production, and it tends to increase production cost. It will take time to ramp up yield learning based on an understanding of defect types in such mixed-technology SOCs. Therefore, defect-tolerant designs are especially important for the emerging marketplace.

Yield enhancement through space redundancy and reconfiguration has been successfully applied to memories, processor arrays (PAs) and field-programmable gate arrays (FPGAs) [5, 6]. The success of these techniques can be attributed to the high regularity of memories, PAs and FPGAs, and the ease with which they can be tested and reconfigured to avoid faulty elements. Digital microfluidics-based biochips are also amenable to redundancy-based yield enhancement. As in the case of memories, they contain regular arrays of small elements, and the elements are simple and identical. Similar to FPGAs, reconfigurability is an inherent property of these devices.

In this paper, we propose a scheme for incorporating defect tolerance in the design of digital microfluidics-based biochips. While spare rows/columns around a mesh-connected array are often used in fault-tolerant processor arrays and FPGAs [6], the property of "fluidic locality" prevents the application of this simple redundancy technique to microfluidic biochips. Due to the absence of programmable interconnects such as switches between microfluidic cells, a droplet is only able to move directly to the adjacent cells. Thus, a faulty cell can only be replaced by its physically-adjacent cells. Consequently, a complicated "shifted replacement" process is required to utilize the spare cells located in the boundary row/column; this results in an unacceptable increase in the reconfiguration cost.

We propose an interstitial redundancy approach to address the above problem. In this approach, spare cells are placed in the interstitial sites within the microfluidic array such that a spare cell can functionally replace any faulty cells that are physically adjacent to it. This defect tolerance method owes its effectiveness to the high utilization of local reconfiguration. We apply this space redundancy technique to a new biochip design with hexagonal electrodes. Microfluidic biochips with different levels of redundancy can be designed to target given yield levels and manufacturing processes. We introduce a metric called "effective yield" to evaluate the yield enhancement provided by these defect-tolerant designs. A set of real-life bioassays, i.e., multiplexed *in-vitro* diagnostics on human physiological fluids, is used to evaluate the proposed method. Simulation results show that the yield of a digital microfluidics-based biochip can be significantly increased with the addition of interstitial redundancy and the use of local reconfiguration.

The organization of the remainder of the paper is as follows. In Section 2, we discuss related prior work. Section 3 presents an overview of digital microfluidics-based biochips. It also introduces a new design based on hexagonal electrodes. Section 4 discusses manufacturing defects and briefly presents a unified test methodology. Reconfiguration techniques for microfluidic biochips are also presented. In Section 5, we introduce various defect-tolerant designs with different levels of redundancy. The defect tolerance of these designs is evaluated in Section 6. In Section 7, multiplexed *in-vitro* diagnostics on human physiological fluids is used to

---

* This research was supported by the National Science Foundation under grant number IIS-0312352

evaluate the proposed yield improvement methodology. Finally, conclusions are drawn in Section 8.

## 2. Related Prior Work

Defect tolerance techniques have been successfully used for memory chips since the late 1970's [5]. In contrast to memory arrays, few logic circuits have been designed with built-in redundancy. The absence of regularity in these circuits usually leads to high overhead. Regular circuits, such as processor arrays and FPGAs require less redundancy; a number of defect tolerance techniques have been proposed to enhance their yield [6, 7].

Microelectromechanical systems (MEMS) is a relatively young field compared to integrated circuits. It employs micromachining techniques, such as surface micromachining and bulk micromachining, in the fabrication process [8]. These processes are less mature than standard CMOS manufacturing processes. As a result, the yield for MEMS devices is often less than that for integrated circuits. Attempts have been made in recent years to make MEMS defect-tolerant. For example, design-for-manufacturing has been incorporated in the design process for MEMS [9].

Microfluidics differs from MEMS in the underlying energy domains and in the working principles. Hence, defect tolerance techniques for MEMS cannot be directly applied to microfluidic biochips. Recent work has focused on a fault classification and a unified test methodology for digital microfluidics-based biochips [10, 11]. Faults are classified as either manufacturing or operational, and techniques have been developed to detect these faults by electrostatically controlling and tracking the droplet motion. This cost-effective test methodology facilitates defect tolerance for digital microfluidics-based biochips.

## 3. Digital Microfluidics-Based Biochips

Digital microfluidics-based biochips manipulate nanoliter-volume droplets using electrowetting. The basic cell in a digital microfluidics-based biochip consists of two parallel glass plates, shown in Figure 1(a). The bottom plate contains a patterned array of individually controllable electrodes, and the top plate contains a ground electrode. The droplets containing biomedical samples are sandwiched between these two plates, and surrounded by a filler medium such as silicone oil. In addition, the droplets are insulated from the electrode array by Parylene C (~800 nm), and a thin layer of hydrophobic Teflon AF 1600 (~50 nm) is coated onto the top and bottom plates to decrease the wettability of the surface.

Electrowetting is the basic principle of microdroplet transportation wherein the interfacial tension of a droplet is modulated with an electric field. A control voltage is applied to an electrode adjacent to the droplet and at the same time the electrode just under the droplet is deactivated. Thus, an accumulation of charge in the droplet/insulator interface over the activated electrode results in a surface tension gradient, which consequently causes the transportation of the droplet. By varying the electrical potential along a linear array of electrodes, nanoliter-volume droplets can be transported along this line of electrodes. The velocity of the droplet can be controlled by adjusting the control voltage (0 ~ 90 V), and droplets have been observed with velocities up to 20cm/s [12]. Furthermore, based on this principle, microdroplets can be transported freely to any location on a two-dimensional array without the need for pumps and valves. The configurations of the microfluidic array are programmed into a microcontroller that controls the voltages of electrodes in the array.

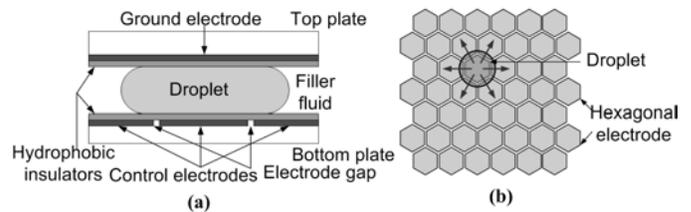

Figure 1: (a) Basic cell used in a digital microfluidics-based biochip; (b) Digital microfluidics-based biochip with hexagonal electrodes.

In the latest generation of microfluidic biochips, hexagonal electrodes are being used to replace the conventional square electrodes design; this close-packed design is expected to increase the effectiveness of droplet transportation in a 2-D array. The top view of a microfluidic array with hexagonal electrodes is shown in Figure 1(b). A droplet can be moved to an adjacent cell in six possible directions. In this paper, we attempt to make this hexagonal array defect-tolerant through space redundancy and local reconfiguration.

## 4. Manufacturing Defects and Reconfiguration Techniques

Digital microfluidics-based biochips are fabricated using standard microfabrication techniques, the details of which are described in [13]. Microfluidic biochips exhibit behavior resembling that of analog and mixed-signal devices. Therefore, we can classify faults caused by the manufacturing defects as being either catastrophic or parametric, along the lines of fault classification for analog circuits [14]. Catastrophic (hard) faults lead to a complete malfunction of the system, while parametric (soft) faults cause a deviation in system performance. A parametric fault is detectable only if this deviation exceeds the tolerance in system performance. However, due to the underlying mixed technology and multiple energy domains, microfluidic biochips exhibit failure mechanisms and defects that are significantly different from failure modes in integrated circuits.

Catastrophic faults may be caused by the following manufacturing defects:

- *Dielectric breakdown*: The breakdown of the dielectric at high voltage levels creates a short between the droplet and the electrode. When this happens, the droplet undergoes electrolysis preventing further transportation.
- *Short between adjacent electrodes*: If a short occurs between two adjacent electrodes, the two electrodes shorted effectively form one longer electrode. When a droplet resides on this electrode, it cannot overlap its adjacent electrodes. As a result, the actuation of the droplet can no longer be achieved.
- *Open in the metal connection between the electrode and the control source*: This defect results in a failure in activating the electrode for transport.

Manufacturing defects that cause parametric faults include geometrical parameter deviations. The deviation in insulator thickness, electrode length and height between parallel plates may exceed their tolerance value during fabrication. Reconfiguration can be employed not only after the detection of catastrophic faults, but also after the detection of parametric faults that cause significant performance degradation

To test a biochip, stimuli droplets containing the normal conducting fluid (e.g., KCL solution) from the droplet source are transported through the array (traversing the cells) to detect the faulty cells. Reconfiguration techniques can be used to bypass faulty cells and increase yield. These reconfiguration approaches can be divided into two categories. The first category consists of techniques that do not add space redundancy, i.e., spare cells, to the microfluidic array. Instead, they attempt to tolerate the defect by using fault-free unused cells. In order to achieve satisfactory yield using this method, fault tolerance must be considered in the design procedure, e.g., in the placement of microfluidic modules in the array. Consequently, it leads to an increase in design complexity. The second category of reconfiguration techniques is application-independent because it incorporates physical redundancy in the microfluidic array. Built-in spare cells can be utilized to replace a defective cell. Since a faulty cell is replaced by a neighboring spare cell, these techniques are also referred to as *local reconfiguration*.

## 5. Defect-Tolerant Designs with Different Redundancy Levels

There are several ways to include spare cells in a defect-tolerant microfluidic array. The first approach is to include spare rows/columns around the microfluidic array. This is a common redundancy technique for PAs and FPGAs. However, in contrast to these electronic arrays with well-defined roles of logic blocks and interconnect, cells in a microfluidic array can be used for storage, transport or other functional operations on droplets. Due to the absence of separate interconnect entities, droplets can only move to physically-adjacent cells. This property is referred to as *microfluidic locality*. Consequently, the functionality of a faulty cell can only be assumed by its physically-neighboring cells in the array. Microfluidic locality limits the reconfiguration capabilities of the spare rows/columns if they are not adjacent to the faulty cell. In order to utilize the spare cell in the boundary rows/columns, a series of replacement, referred to as "shifted replacement" is required. In *shifted replacement*, each faulty cell is replaced by one of its fault-free adjacent cells, which is in turn replaced by one of its adjacent cells, and so on, until a spare cell from the boundary is incorporated in the reconfigured structure. In many cases, this shifted replacement procedure will not only involve the faulty module, but it will also require the reconfiguration of fault-free modules. Therefore, it significantly increases the complexity of the reconfiguration. Figure 2 shows an example of a microfluidic array with a single spare row. If one cell in Module 1 is faulty, Module 1 can be only relocated to bypass the faulty cell, while other modules remain unchanged; see Figure 2(b). However, if there is one faulty cell in Module 3, the shifted replacement of Module 3 causes the reconfiguration of Module 2 even

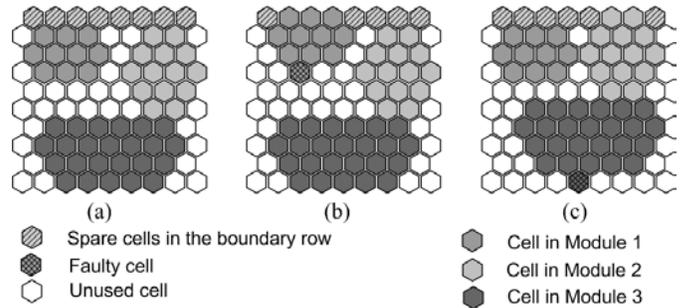

Figure 2: Example of a microfluidic array with one single spare row.

though it is fault-free; see Figure 2(c).

In order to address the problems resulting from microfluidic locality, a new space redundancy approach, termed *interstitial redundancy* [7], is proposed in this paper. In this approach, spare cells are located in the interstitial sites within the microfluidic array such that each spare cell is able to functionally replace any one of the primary cells adjacent to it. In contrast to redundancy based on boundary spare rows/columns, interstitial redundancy offers a simple reconfiguration scheme that effectively utilizes local reconfiguration. We apply interstitial redundancy to a digital microfluidics-based biochip with hexagonal electrodes. Such defect-tolerant microfluidic arrays can incorporate different levels of redundancy depending on the number and location of spare cells. We next introduce some key definitions.

*Definition* 1: A defect-tolerant design for a digital microfluidics-based biochip, denoted $DTMB(s, p)$, has interstitial spare cells such that each non-boundary primary cell can be replaced by any one of $s$ spare cells, and each spare cell can be used to replace any one of $p$ primary cells.

*Definition* 2: The redundancy ratio ($RR$) for a defect-tolerant microfluidic array with interstitial redundancy is the ratio of the number of spare cells in the array to the number of primary cells. Clearly, for a $DTMB(s, p)$ array of large size, $RR \approx s/p$.

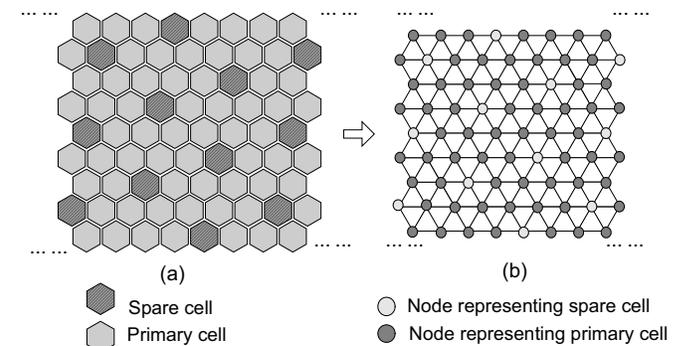

Figure 3: Top view and graph model of $DTMB(1, 6)$.

A $DTMB(1, 6)$ design is shown in Figure 3(a). A corresponding graph model, derived from the array, is shown in Figure 3(b). Black nodes in the graph represent the primary cells in the microfluidic biochip, while white nodes denote spare cells. An edge between two nodes indicates that the two cells represented by these nodes are physically adjacent in the array. Each primary cell is adjacent to only one spare cell, and every spare cell is adjacent to six primary cells. Therefore, the redundancy ratio for this array approaches 0.1667 as the array size increases.

Other defect-tolerant array designs, e.g., *DTMB*(2, 6), *DTMB*(3, 6) and *DTMB*(4, 4), are shown in Figures 4 ÷ 6. The redundancy ratios of the different designs are listed in Table 1.

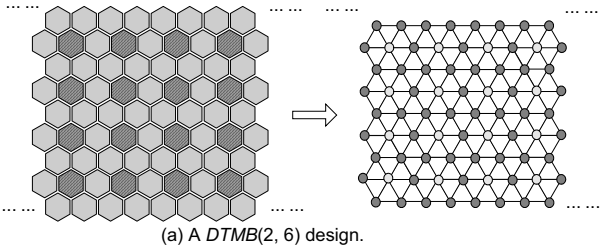
(a) A *DTMB*(2, 6) design.

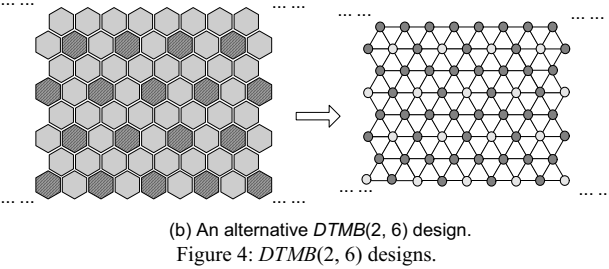
(b) An alternative *DTMB*(2, 6) design.
Figure 4: *DTMB*(2, 6) designs.

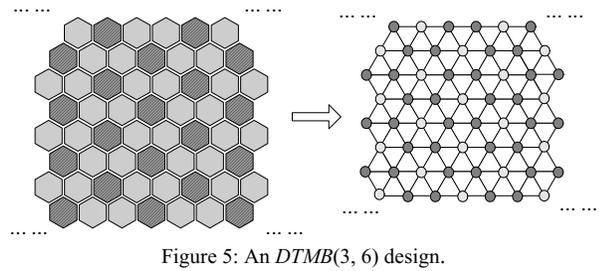
Figure 5: An *DTMB*(3, 6) design.

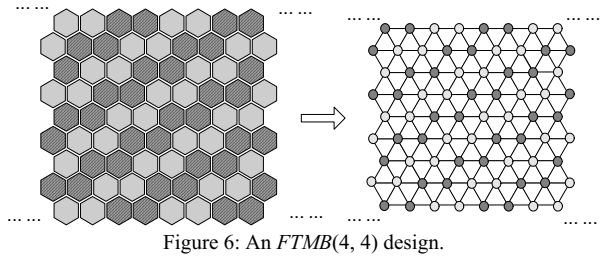
Figure 6: An *FTMB*(4, 4) design.

Table 1: Redundancy ratios for the different defect-tolerant architecture.

| Design | *DTMB*(1, 6) | *DTMB*(2, 6) | *DTMB*(3, 6) | *DTMB*(4, 4) |
|---|---|---|---|---|
| RR | 0.1667 | 0.3333 | 0.5000 | 1.0000 |

## 6. Estimation of Yield Enhancement

The effectiveness of various defect-tolerant designs can be determined by estimating their enhanced yields. The yield analysis in this paper is based on the following assumption.

*Assumption*: Each single cell in the microfluidic array including each primary and spare cell, has the same defect probability $q$. Moreover, the failures of the cells are independent. Let $p = 1 - q$ denote the survival probability.

Note that the assumption of equal survival probabilities is reasonable since each cell in the microfluidic array has the same structure. In addition, the assumption of independent failures is valid for random and small spot defects, which result from imperfect materials and from undesirable chemical and airborne particles.

Based on these assumptions, the yield for a defect-tolerant design can be obtained in terms of $p$. We use both analytical modeling and Monte-Carlo simulation.

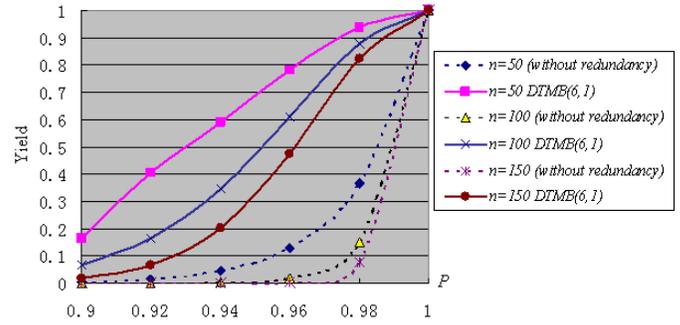
Figure 7: Yield for *DTMB*(1,6).

Since each primary cell is physically adjacent to only one spare cell in *DTMB*(1, 6), the spare assignment to a faulty cell is straightforward. Thus, its yield can be easily obtained analytically. We can view *DTMB*(1, 6) as a composition of identical clusters that consist of one spare cell and six primary cells surrounding the spare cell. The yield $Y_c$ of any cluster in *DTMB*(1, 6) is determined by the likelihood of having at most one failed cell among these seven cells, i.e., $Y_c = p^7 + 7p^6(1-p)$. A biochip with $n$ primary cells can be approximately divided into $n/6$ clusters. Since the cluster failures are independent, the yield $Y$ for this design is given by $Y = Y_c^{n/6} = \left(p^7 + 7p^6(1-p)\right)^{n/6}$. Figure 7 shows the yield for *DTMB*(1, 6) for different values of $p$ and $n$, and compares it to the yield for a biochip without redundancy. Clearly, interstitial redundancy improves the yield of the microfluidic biochip.

For the defect-tolerant design with a higher level of redundancy such as *DTMB*(2, 6), *DTMB*(3, 6) and *DTMB*(4, 4), it is hard to develop an analytical model to determine the yield because their spare assignments are not straightforward. We therefore address this problem using Monte-Carlo simulation. During each run of the simulation, the cells in the microfluidic array, including both primary and spare cells, are randomly chosen to fail with probability $p$. We then check if these defects can be tolerated via local reconfiguration based on the interstitial spare cells. This checking procedure is based on a graph matching approach as described below.

We develop a bipartite graph model to represent the relationship between faulty and spare cells in the microfluidic array. A *bipartite graph BG*($A$, $B$, $E$) is a graph whose nodes can be partitioned into two sets, $A$ and $B$, and each edge in

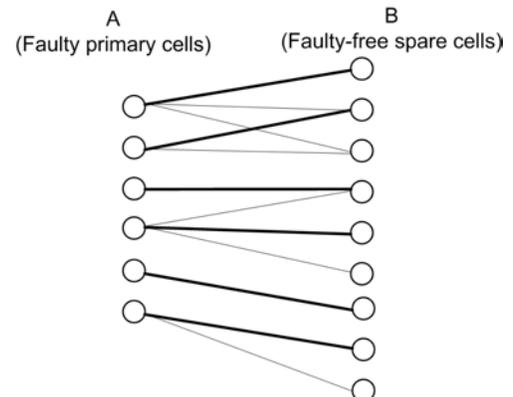
Figure 8: Example of maximal bipartite matching model.

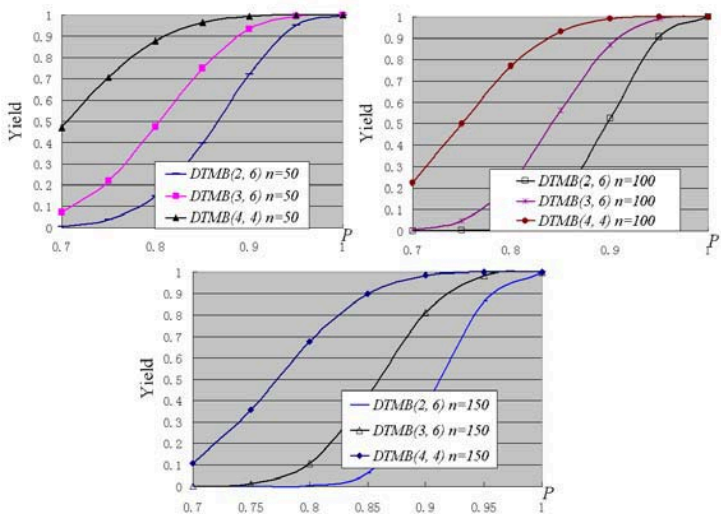

Figure 9: Yield estimation for *DTMB*(2,6), *DTMB*(3,6) and *DTMB*(4,4).

*E* has one node in *A* and one node in *B* [15]. In our model, nodes in *A* represent the faulty primary cells in the microfluidic array while nodes in *B* denote the fault-free spare cells. An edge exists from a node *a* in *A* to a node *b* in *B* if and only if the faulty primary cell represented by *a* is physically adjacent to the spare cell represented by *b*. An example is shown in Figure 8. A maximal matching for this bipartite graph can be obtained using well-known techniques [16]. If this maximal matching covers all nodes in *A*, it implies that all faulty cells can be replaced by their adjacent fault-free spare cells through local reconfiguration. Otherwise, this microfluidic biochip cannot be reconfigured. After 10000 simulation runs, the yield of this microfluidic array is determined from the proportion of successful reconfigurations. The simulation results for *DTMB*(2, 6), *DTMB*(3, 6) and *DTMB*(4, 4) are shown in Figure 9, where *n* is the number of primary cells.

From Figure 9, it is clear that a higher level of redundancy leads to a higher yield. However, adding more redundant cells increases the array area and thereby manufacturing cost. To measure yield enhancement relative to the increased array size, we define the effective yield *EY* as: $EY = Y \times (n/N) = Y/(1+RR)$, where *n* is the number of primary cells, and *N* is the total number of cells in the microfluidic array. The parameter *EY* represents the tradeoff between yield enhancement and increase in manufacturing cost. The variation of *EY* with *p* for different redundancy levels is shown in Figure 10. The number of primary cells is set to 100. As expected, the results show that a microfluidic structure with the higher level of redundancy, such as *DTMB*(4, 4), is suitable for small values of *p*. On the other hand, a lower level of redundancy, such as *DTMB*(1, 6) or *DTMB*(2, 6), should be used when *p* is relatively high.

## 7. Example: Multiplexed *in-vitro* Diagnostics

In this section, we evaluate the proposed defect-tolerant design by applying it to a digital microfluidics-based biochip used for multiplexed biomedical assays. The *in-vitro* measurement of glucose and other metabolites, such as lactate, glutamate and pyruvate, in human physiological fluids plays a critical role in clinical diagnosis of metabolic disorders. For example, the change of regular metabolic parameters in the patient's blood can signal organ damage or dysfunction prior to observable microscopic cellular damages or other symptoms. Recently, the feasibility of performing a colorimetric enzyme-kinetic glucose assay on a digital microfluidic biochip has been successfully demonstrated in experiments [17].

The glucose assay performed on the biochip is based on Trinder's reaction, a colorimetric enzyme-based method [18]. The enzymatic reactions involved in the assay are:

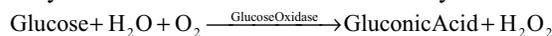
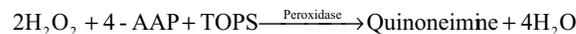

In the presence of Glucose oxidase, glucose can be enzymatically oxidized to gluconic acid and hydrogen peroxide. Then, in the presence of peroxidase, the hydrogen peroxide reacts with 4-amino antipyrine (4-AAP) and N-ethyl-N-sulfopropyl-m-toluidine (TOPS) to form violet-colored quinoneimine, which has an absorbance peak at 545nm. Based on Trinder's reaction, a complete glucose assay can be performed following three steps, namely, transportation, mixing and optical detection. A sample droplet containing glucose and a reagent droplet containing glucose oxidase, peroxidase, 4-AAP and TOPS, are dispensed into the microfluidic array from their respective droplet sources. They are then transported towards a mixer where the sample and the reagent droplets are mixed. The mixed droplet is transported onto a transparent electrode to enable observation of the absorbance of the products of the enzymatic reaction. Absorbance measurements are performed with a green LED and a photodiode. The glucose concentration can be measured from the absorbance, which is related to the concentration of colored quinoneimine in the droplet. Besides glucose assays, other metabolites such as lactate, glutamate, and pyruvate have been detected in a digital microfluidic biochip recently. In addition to current metabolites, a number of reagents can be added onto a chip to enable a multiplexed *in-vitro* diagnostics platform on different human physiological fluids. Figure 11 shows a recently fabricated microfluidic biochip used for multiplexed biomedical assays [17]. In this biochip with square electrodes, SAMPLE1 and SAMPLE2 contain glucose and REAGENT1 and REAGENT2 contain the reagents.

In this first design and demonstration, only cells used for the bioassays were fabricated; no spare cells were included in

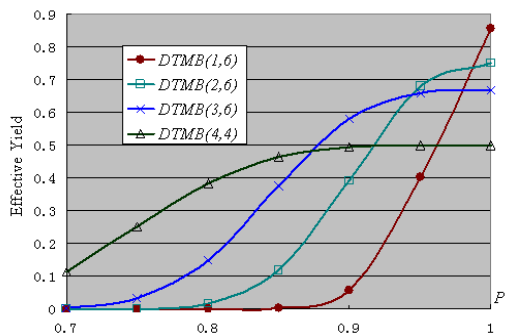

Figure 10: Effective yield for different levels of redundancy.

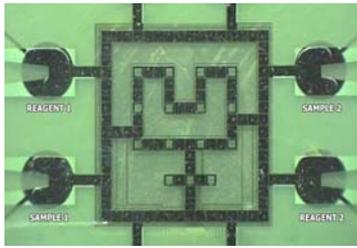

Figure 11: Fabricated biochip used for multiplexed bioassays.

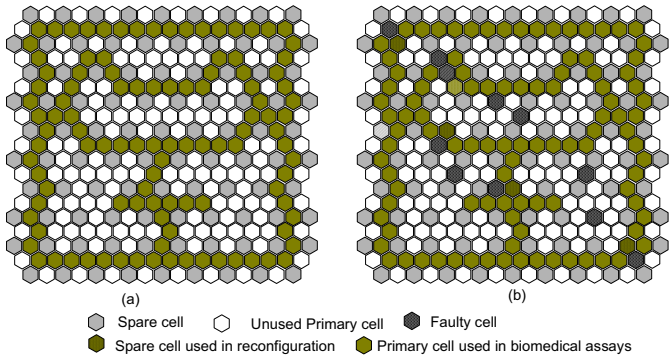

(a)          (b)

☐ Spare cell    ☐ Unused Primary cell    ■ Faulty cell
■ Spare cell used in reconfiguration    ■ Primary cell used in biomedical assays the array. Thus, even if one arbitrary cell in this biochip becomes faulty due to a manufacturing defect, this failure cannot be avoided by reconfiguration, and the fabricated biochip has to be discarded. Consequently, the yield for this biochip design is very low. It is only 0.3378 even if the survival probability of a single cell is as high as 0.99. Such low yield makes the first biochip design unsuitable for future mass fabrication and use in clinic diagnostics.

Figure 12: (a) A defect-tolerant design based on $DTMB(2, 6)$. (b) An example of local reconfiguration.

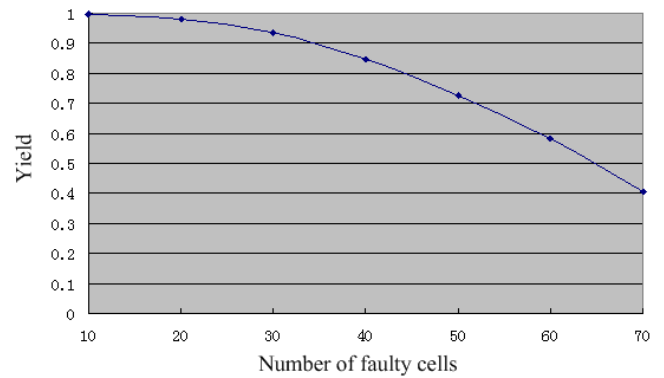

Figure 13: Yield estimation for the $DTMB(2,6)$-based design in the presence of multiple defects.

In order to improve the yield, we use a defect-tolerant design with interstitial as described in Section 6. To facilitate the comparison, the topological structure of primary cells in the first design is directly mapped to a $DTMB(2, 6)$ design. The new defect-tolerant design has the same number of primary cells used for multiplexed biomedical assays as the original design; see Figure 12(a). There are 252 primary cells (108 of them used in assays) and 91 spare cells in this defect-tolerant biochip.

To analyze the improvement in yield, we randomly introduce $m$ cell failures, and then apply local reconfiguration to avoid them. The yield in the presence of $m$ failures is obtained through Monte-Carlo simulation. The yield for the different values of $m$ is shown in Figure 13. For up to 35 faults, the redundant design can provide a yield of at least 0.90. An example of successful reconfiguration in the presence of 10 faulty cells is shown in Figure 12(b).

## 8. Conclusions

We have presented a yield enhancement technique for digital microfluidics-based biochips. This technique relies on (i) space redundancy, whereby spare cells are placed in the interstitial sites of the microfluidic array, and (ii) local reconfiguration, in which spare cells replace the neighboring faulty cells. The defect-tolerant design has been evaluated for a set of real-life bioassays. Low yield, which is expected to be a consequence of increased area and density of biochips, will be a deterrent to high-volume production and it will increase production cost. The proposed defect-tolerant design approach will therefore be especially useful for the emerging marketplace.